%% file: eurosys27_oxidize_technicalreport.tex
\documentclass[sigplan,11pt,nonacm]{acmart}

\def\NoMinted{}

\input{header/packages}
\input{header/config}

\input{header/commands}

\input{header/boiler}
\copyrightyear{2027}
\acmYear{2027}
\setcopyright{cc}
\setcctype{by}

\makeatletter
\AtBeginEnvironment{Verbatim}{\let\@vspace\@vspace@orig\let\@vspacer\@vspacer@orig}
\makeatother

\makeatletter
\def\@mkabstract{\bgroup
  \ifx\@abstract\@lempty\else
    \begin{center}\small\bfseries\abstractname\end{center}\vskip -1ex%
    \begin{quote}\small\ignorespaces\@abstract\par\end{quote}%
  \fi\egroup}
\renewcommand\tableofcontents{%
  {\noindent\Large\bfseries\contentsname\par}%
  {\let\addcontentsline\@gobblethree\@starttoc{toc}{}}}
\def\l@section{\@tocline{1}{6pt}{0pt}{1.5em}{\bfseries}}
\makeatother

\begin{document}

\title[Lifting the Preprocessor with \sys{} (Technical Report)]{Lifting the Preprocessor with \sys{}: Structure-Preserving C-to-Rust Translation\texorpdfstring{\\}{ }(Technical Report)}

\input{header/authors}

\renewcommand{\abstractname}{About this document}
\begin{abstract}
This technical report contains material accompanying our work with
the same title published at EuroSys'27
(\url{https://doi.org/10.1145/3842654.3848533}). The paper is self-contained;
this report collects the additional results, extended tables, and
secondary experiments that the paper, referred to here as ``the
main paper'', cites in passing.
\end{abstract}

\onecolumn
\makeatletter
{\let\addcontentsline\@gobblethree
 \renewcommand\twocolumn[1][]{#1\par}%
 \maketitle}
\makeatother

\vskip\bigskipamount
\tableofcontents
\clearpage

\renewcommand{\thesection}{\Alph{section}}
\setcounter{section}{0}

\input{body/supplementary_implementation}

\input{body/supplementary_macro_typing}

\input{body/supplementary_codebases}

\input{body/supplementary_population}

\input{body/supplementary_safety}

\input{body/supplementary_overhead}

\input{body/supplementary_prompts}

\bibliographystyle{ACM-Reference-Format}
\bibliography{body/references}

\end{document}

%% file: header/packages.tex
\usepackage{color}
\usepackage{xspace}
\usepackage{enumitem}
\usepackage{float}
\ifdefined\NoMinted
  \usepackage{fvextra}
\else
  \usepackage[newfloat]{minted}
\fi
\usepackage{subcaption}
\usepackage{graphicx}
\usepackage{pifont}
\usepackage[dvipsnames]{xcolor}
\usepackage{multirow}
\usepackage{multicol}
\usepackage{makecell}
\usepackage{tikz}

%% file: header/config.tex
\usetikzlibrary{positioning, arrows.meta}

\ifdefined\NoMinted
\makeatletter
\AtBeginEnvironment{Verbatim}{\let\@vspace\@vspace@orig\let\@vspacer\@vspacer@orig}
\makeatother
\else
\setminted{
  tabsize=4,
  fontsize=\scriptsize,
  linenos,
  numbersep=4pt,
  xleftmargin=14pt,
}
\fi

\makeatletter
\input{t1zi4.fd}
\input{ts1zi4.fd}
\makeatother
\DeclareFontShape{T1}{zi4}{m}{it}{<->ssub * zi4/m/n}{}
\DeclareFontShape{TS1}{zi4}{m}{it}{<->ssub * zi4/m/n}{}

\makeatletter
\g@addto@macro\UrlBreaks{\do\-\do\.\do\_}
\makeatother

%% file: header/commands.tex
\newcommand{\sys}{Oxidize}

\definecolor{FOOBAR}{HTML}{F00BA2}

\ifdefined\NoMinted
  \newcommand{\inlinec}{\Verb}
  \newcommand{\inlinerust}{\Verb}
\else
  \newcommand{\inlinec}{\mintinline{c}}
  \newcommand{\inlinerust}{\mintinline{rust}}
\fi


%% file: header/boiler.tex
\AtBeginDocument{%
  }

%% file: header/authors.tex
\author{Robbe De Greef}
\orcid{0009-0000-1477-2616}
\affiliation{%
  \institution{Vrije Universiteit Brussel}
  \city{Brussels}
  \country{Belgium}
}
\email{robbe.de.greef@vub.be}

\author{Th\'eo Engels}
\orcid{0009-0009-5405-186X}
\affiliation{%
  \institution{Royal Military Academy}
  \city{Brussels}
  \country{Belgium}
}
\email{theo.engels@mil.be}

\author{Felix Van den Broucke}
\orcid{0009-0004-1674-7878}
\affiliation{%
  \institution{Vrije Universiteit Brussel}
  \city{Brussels}
  \country{Belgium}
}
\email{felix.van.den.broucke@vub.be}

\author{Ken Hasselmann}
\orcid{0000-0002-8196-9889}
\affiliation{%
  \institution{Royal Military Academy}
  \city{Brussels}
  \country{Belgium}
}
\email{ken.hasselmann@mil.be}

\author{Antonio Paolillo}
\orcid{0000-0001-6608-6562}
\affiliation{%
  \institution{Vrije Universiteit Brussel}
  \city{Brussels}
  \country{Belgium}
}
\email{antonio.paolillo@vub.be}

\renewcommand{\shortauthors}{R. De Greef, T. Engels, F. Van den Broucke, K. Hasselmann, A. Paolillo}

%% file: body/supplementary_implementation.tex
\section{\sys{} Implementation Details}
\label{sec:supp-implementation}

This section expands the architectural overview given in
\S4 of the main paper. It collects the
design rationale (three-phase architecture, incremental passes,
IR shape) and the front/middle/back-end implementation choices
that the main paper summarises in a single paragraph for space
reasons.

\paragraph{Three-phase architecture.}
Like modern compilers~\cite{lattner2021mlir,dragonbook}, \sys{}
adopts a front, middle, and back end design with an intermediate
representation (IR) at the core. The front end parses C code and
translates it into an annotated IR. The middle end is currently
empty, reserved for downstream extensions
(see \S\ref{sec:supp-safety-roadmap}). The back end rewrites the
IR into Rust AST form and generates source code. While such a
structure is standard in compilers, source-to-source translators
are usually built for a single language pair
(e.g., Emscripten~\cite{emscripten}, Dart2js~\cite{dart2js}); some
walk the source AST and print target code directly, with no
intermediate representation at all (Pyjs~\cite{pyjs_translator}). For \sys{}, the explicit IR boundary is
crucial: it lets macro-related transformations stage
systematically and makes the system extensible to other source
or target languages.

\paragraph{Incremental passes.}
The distinctive feature of \sys{} is its use of small,
incremental passes over the IR. This approach, inspired by
nanopass~\cite{sarkar2004nanopass} and MLIR~\cite{lattner2021mlir},
decomposes translation into fine-grained transformations, each
with a narrow responsibility. In the front end, passes contract
macro definitions into IR nodes
(\S4.1 of the main paper),
resolve nested macro definitions via invoke-expand
(\S4.2), check parameter hygiene
(\S4.3), and type macro arguments
(\S4.4). In the back end, passes convert
contracted macros into Rust syntax or generate helper macros
where direct IR constructs are insufficient. The design enables
experimentation with alternative algorithms and incremental
validation of each transformation; it also makes \sys{} amenable
to extension---new passes can be added without rearchitecting
the whole translator.

\paragraph{Intermediate representation.}
The IR is a high-level, AST-like representation that bridges C
and Rust. Unlike LLVM IR, which targets machine code, \sys{}'s
IR preserves source-level constructs relevant to maintainability.
Two principles guided its design. First, the IR mirrors a Rust
AST, with control flow expressed as expressions, items instead of declarations, and Rust-like types, but
admits C semantics that Rust lacks (e.g., assignments that yield
a value, raw-pointer arithmetic). Second, the IR is parameterised
by a \emph{marker} type: nodes may carry an extension object,
which the front end instantiates with the C-only constructs Rust
has no counterpart for (Table~\ref{tab:supp-ir-nodes}). Passes
lower these markers one kind at a time. When none is left, the
tree is rebuilt at the marker-free instantiation of the same IR,
the one every later stage is typed against. Nothing in the tree
changes there, only its type: from that point on it is the type
system and not a convention that keeps C-only constructs out of
the back end, and a marker that somehow survived is caught at the
conversion. Because the IR keeps macro definitions and
invocations as first-class nodes, \sys{} avoids the information
loss of preprocessor-based approaches.

\paragraph{IR node taxonomy.}
Table~\ref{tab:supp-ir-nodes} lists the node kinds, about 70 in
total (operator enumerations not counted). Three groups matter
for macro translation. \emph{Macro definitions} store the
replacement body as a single IR expression, the parameters (each
an expression, an identifier, or a type), and the body's
\emph{external references}: every free identifier together with
the scope level at which it was bound at the definition site,
which the hygiene check (\S4.3 of the main paper) compares against
the use site. \emph{Macro invocations} are ordinary expression
nodes whose arguments are expressions, types, identifiers, or
tokens; type and token arguments carry the parameters added by
macro typing (\S4.4 of the main paper). \emph{C-only markers}
represent the constructs with no Rust counterpart; each is
eliminated by a dedicated front-end pass (Table~\ref{tab:supp-ir-nodes},
last row).

\begin{table*}[t]
  \centering
  \small
  \begin{tabular}{@{}l r p{0.68\linewidth}@{}}
    \toprule
    Category & Kinds & Node kinds \\
    \midrule
    Items & 12 & function; macro definition; include (external, regular, or generated); module; struct, union, and enum definitions; constant; static; type alias; trait; impl (traits and impls host the generated helper traits) \\
    Statements & 4 & expression; void expression; local declaration; nested item \\
    Expressions & 29 & control flow: block (with \texttt{unsafe} flag and label), if, while, for, loop, break, continue, return, match, unreachable;
      operators: binary (18 operators), unary (dereference, not, negation, bitwise not);
      memory: address-of (reference or raw borrow), raw-pointer arithmetic (add, subtract, offset), pointer indexing, cast, \texttt{sizeof};
      data: literal (7 kinds), path, struct and union initialisers, array initialiser, member access, method call, call, default value;
      C-style assignment and compound assignment (both yield a value);
      macro invocation \\
    Types & 9 & primitive (integer, float, string, C string, byte string, char, void, bool); function (fixed or variadic parameter list); named object; reference; raw pointer; alias; array; generic; never \\
    Patterns & 4 & or-pattern (in matches, e.g.\ \texttt{p | q}); literal; identifier
      binding; wildcard (\texttt{\_}) \\
    C-only markers & 10 & do-while (lowered by the do-while conversion); switch, case, default, label, goto (lowered by control-flow restructuring); pre- and post-increment/decrement (lowered to helper macros); unary plus (dropped during macro-body parsing); pending macro definition (resolved by the macro-validity pass into a definition or an expansion) \\
    \bottomrule
  \end{tabular}
  \caption{Node kinds of \sys{}'s IR. C-only markers exist only in
  the front end and must all be lowered before the IR is unmarked.}
  \label{tab:supp-ir-nodes}
\end{table*}

\paragraph{Pass scheduling.}
Every pass is a function from IR to IR over one translation unit.
Passes are composed in a fixed order written in the driver: there
is no pass manager, dependency graph, or fixpoint iteration, and
the few ordering constraints are local and documented in the code
(e.g., aliases are corrected before statics are
default-initialised, because the latter queries types; generated
helper modules are unpacked before the module tree is built). Each
translation unit goes through six stages:
\begin{enumerate}
\item \emph{Contracting translation.} The libclang AST is converted
  into the marked IR. Before converting the program, macro
  definitions are parsed and classified: the macro dependency
  graph is resolved leaves first, a macro stays \emph{invoke} only
  if its body parses as a C expression once its dependencies are
  resolved, and a cycle forces \emph{expand} (invoke-expand, \S4.2
  of the main paper). While converting expressions, any
  expression whose source range matches a recorded macro expansion
  is replaced by an invocation node, and the expansion is kept for
  a later rollback (macro contraction, \S4.1). Variadic macros and
  macros using \texttt{\#} or \texttt{\#\#} are left to expansion.
\item \emph{Front-end passes on the marked IR (20).} Lowering of
  C-only constructs (do-while, switch/goto/labels, increment and
  decrement operators); C semantic corrections (array decay,
  enumeration accesses, type aliases, static and aggregate
  initialisers, function-pointer dereferences, a default
  \texttt{return} for non-void functions that lack one, variadic
  parameters); and the
  macro passes: the validity and hygiene check, which rolls an
  invocation back to its stored expansion when it calls an invalid
  macro or when a free identifier of the body resolves to a
  different scope at the use site (\S4.3), followed by macro typing,
  which adds type parameters to macro definitions and rewrites
  operators inside macro bodies into helper-trait calls (\S4.4).
\item \emph{Unmarking,} which checks that no marker remains.
\item \emph{Front-end passes on the unmarked IR (8).} Pointer
  arithmetic made explicit as raw-pointer operations; C's usual
  arithmetic conversions made explicit as casts; array casts and
  primitive sizes; top-level \texttt{unsafe} blocks; negation of
  unsigned values; renaming of Rust keywords; non-constant
  \texttt{sizeof}.
\item \emph{Middle end (0).} Currently empty: this is where the
  safety-refinement passes of \S\ref{sec:supp-safety-roadmap} would
  run.
\item \emph{Back-end passes before emission (7).} Function pointers
  as \texttt{Option<fn>}; casts inside macros routed through the
  generic cast macro (\S4.4); remaining cast fixes; C assignments
  used as values rewritten to helper-macro calls; renaming of
  shadowed statics and constants; \texttt{va\_arg} mapped to Rust's
  C-variadic API; wrapping negation.
\end{enumerate}
Four whole-program passes then run once over all translation
units: unpacking the generated helper modules, building the Rust
module tree, generating the Rust \texttt{main} trampoline (skipped
in library mode), and renaming module names that clash with Rust
keywords. In total the pipeline has 41 passes: 28 in the front
end, none in the middle end, and 13 in the back end (including the
two below).

\paragraph{Control-flow restructuring.}
Every function body is lowered to basic blocks and restructured
by a stackifier algorithm implemented in a separate crate. Irreducible loops are found
by a recursive search over strongly connected components; a loop
with several entries receives a dispatch node driven by a state
variable, which makes it reducible. Loops, if-scopes, and forward
jumps are then rebuilt from dominator information and a
topological order, the forward jumps as labelled blocks with
\texttt{break}. This is how \sys{} supports \texttt{goto} and C's
\texttt{switch}.

\paragraph{Front-end implementation.}
The current implementation supports only C as input. Parsing is
handled by \texttt{libclang}\footnote{\url{https://clang.llvm.org/}}
with a detailed preprocessing record, which exposes both the
expanded program and every macro expansion with its source range
(essential for macro contraction). Macro bodies are parsed with a
dedicated grammar, since they need not be valid C on their own.

\paragraph{Back-end emission.}
The IR of each translation unit is converted into a
Rust AST using the Ruast library\footnote{\url{https://github.com/mtshiba/ruast}}.
Macro definitions become single-arm \texttt{macro\_rules!}
definitions whose body is the converted expression, with the
parameters prefixed by \texttt{\$}; macro invocations become Rust
macro calls. Two passes on the Rust AST then generate the
multi-arm helper macros: the generic cast macro, with one arm per cast kind
(\S\ref{sec:supp-macro-typing}), and the assignment macros. The helper traits used by macro typing are
emitted as a generated module
(\texttt{oxidize\_generated/c\_macro\_helpers.rs}), by default
only for the operations actually used. Each C file becomes one
Rust file; a module tree with a \texttt{mod.rs} per directory and
a root \texttt{main.rs} (or \texttt{lib.rs} in library mode)
completes the crate, and the result is formatted with
\texttt{rustfmt}. No runtime library is linked: every helper is
emitted as source.

\paragraph{Size.}
\sys{} is about 22,700 lines of Rust (excluding comments and blank
lines): the front end accounts for 9,200 (3,900 of them in passes
and 2,200 in macro parsing), the IR for 5,800, the back end for
2,900, and the control-flow restructuring crate for 4,000.

%% file: body/supplementary_macro_typing.tex
\section{Macro Typing Details}
\label{sec:supp-macro-typing}

This section expands two design points referenced in
\S4 of the main paper (Helper Traits and Cast Macro).

\paragraph{Helper Traits: implementation breadth.}
\sys{} pre-generates the helper-trait module once per
translation. The current build emits a
\inlinerust{Helper<Op><T>} trait for 16 binary operators
(arithmetic, bitwise, shift, comparison) and three unary ones
(dereference, increment/decrement, logical negation), plus an impl
for every pair of supported operand types: signed and unsigned
integers from 8 to 128 bits plus \texttt{size}, \texttt{f32} and
\texttt{f64}, and generic pointer types. Each impl
applies C's implicit conversion
rules and behaves in the way the C counterpart would~\cite[\S6.3.1.8]{c99_standard}.

\paragraph{Helper Traits vs.\ per-call macro generation.}
An alternative design would emit one Rust macro per
operand-type combination
(\inlinec{ADD_i32_i32}, \inlinec{ADD_ptr_i32}, ...). We rejected
this approach because it (i) could potentially require multiple macros to be maintained instead of a single source of truth; and (ii) requires regenerating the macro module each time a downstream developer adds a new type. Macros outside the
pre-generated table fall back to conservative expansion
(see main paper \S3.5).

\paragraph{Cast Macro: dispatch kinds.}
The \texttt{helper\_cast} macro (main paper, \S4.4)
dispatches on six cast kinds, e.g.\ \texttt{regular} (a plain
\texttt{as}), \texttt{multicast} (an array lowered to a pointer), and
\texttt{transmute} (function pointers among others). Additional variants can be added
to support other C-to-Rust cast semantics without touching the
translator core; the macro is intentionally extensible so that
new kinds can be hosted under the same name.

%% file: body/supplementary_codebases.tex
\section{Evaluated Codebase Sources}
\label{sec:supp-codebases}

The nine real-world C codebases used for the macro-preservation
study (\S5.1 in the main paper) are open-source. Two of them, FIGlet and DOOM, also
illustrate the macro-class distribution in \S3.5.
The upstream URLs and the commits pinned for the evaluation are
listed in Table~\ref{tab:supp-codebases}.

\begin{table}[ht]
  \centering
  \small
  \begin{tabular}{@{}l l l r r@{}}
    \toprule
    Codebase            & Upstream URL & Pinned revision (tag or SHA) & C files & kLoC \\
    \midrule
    \texttt{doomgeneric} & \url{https://github.com/ozkl/doomgeneric} & \texttt{fc601639494e} & 84 & 59.0 \\
    \texttt{libpng}      & \url{https://github.com/glennrp/libpng}   & \texttt{v1.6.47} & 24 & 56.5 \\
    \texttt{zlib}        & \url{https://github.com/madler/zlib}      & \texttt{v1.3.2} & 17 & 10.8 \\
    \texttt{bzip2}       & \url{https://gitlab.com/bzip2/bzip2}      & \texttt{6a8690fc8d26} & 9 & 6.9 \\
    \texttt{FIGlet}      & \url{https://github.com/cmatsuoka/figlet} & \texttt{202a0a811065} & 6 & 5.2 \\
    \texttt{lolcat}      & \url{https://github.com/jaseg/lolcat}     & \texttt{acb30b7e1f91} & 2 & 0.6 \\
    \texttt{smaz}        & \url{https://github.com/antirez/smaz}     & \texttt{2f625846a775} & 2 & 0.3 \\
    \texttt{nyancat}     & \url{https://github.com/klange/nyancat}   & \texttt{32fd2eb40332} & 1 & 0.9 \\
    \texttt{sl}          & \url{https://github.com/mtoyoda/sl}       & \texttt{923e7d7ebc5c} & 1 & 0.3 \\
    \bottomrule
  \end{tabular}
  \caption{Upstream sources, pinned revisions and size of the nine codebases
  evaluated in \S5.1 of the main paper (revisions as recorded in the
  artifact's per-codebase configuration; sizes count the C source files
  in each build and their lines of code, 146 files and 141\,kLoC in
  total). Rows are ordered by decreasing number of source files, the
  order used by the figures.}
  \label{tab:supp-codebases}
\end{table}

%% file: body/supplementary_population.tex
\section{Per-Codebase Macro Population}
\label{sec:supp-population}

This section reports the per-codebase macro population resulting
from our macro census, summarised in the main paper (\S5.1,
``Macro population'' paragraph). The
content is descriptive of the input alone (no translator is
involved): it characterises what each codebase declares,
class by class, before any translation choice is made.

\input{figures/tex/fig-macros-population}

The class breakdown is heavily skewed toward the simple end of
the macro spectrum: across the nine codebases, 1519 of the 2862
macro definitions (53\%; 2709 distinct names) are constant literals (object-like macros
whose body is a single integer, string, character, or other
literal). Non-constant object-like macros, whose body names an identifier
(such as \texttt{logical\_gamemission} in Listing~1 of the main
paper; the syntactic class also catches aliases of other
constants), are the second-largest class: 502
definitions (18\%) and 2700 call sites, more than either
function-like class. They are also the object-like class that
translators handle worst: on the intersection of \S5.1, C2Rust
retains 150 of the 283 definitions in scope (53\%), rust-bindgen 167
(59\%), and \sys{} 250 (88\%). The function-like classes still carry
weight, too:
277 function-like-expression macros and 95
function-like-statement macros across the nine codebases. Four
of the nine codebases declare more than 25 function-like macros
each, with \texttt{libpng} declaring 190 (175 expression-like, 15
statement-like); on every codebase
larger than \texttt{FIGlet}, function-like macros are a
substantial fraction of the macro surface.

\paragraph{Name collisions and class attribution.}
A macro name can carry several definitions with different
classes. DOOM, for instance, defines \texttt{R} as an object-like
constant in \texttt{am\_map.c} (\texttt{((8*PLAYERRADIUS)/7)})
and, unrelatedly, as a \texttt{do \{\dots\} while (0)} round
function in \texttt{sha1.c}. The census records call sites per
name, so classifying by name alone would file every call site of
\texttt{R} under whichever definition happened to come last, and
summing per definition would count each site once per definition.
The call-site counts of Figure~\ref{fig:macro-population} and the
retention figure of \S5.1 of the main paper therefore count each
call site once and attribute it (and, for retention, each retained
definition) to the definition visible from the file in question: a
definition in that file wins, and otherwise the remaining
definitions must agree on a class. Conditional
compilation can still leave a genuine disagreement, typically a
name defined empty in one branch and as an expression in
another; such macros are excluded rather than attributed by
guesswork. Two names are affected (\texttt{PNG\_DEPRECATED} in
libpng and \texttt{GZIP} in zlib): 19 of the 14\,419 call sites
of Figure~\ref{fig:macro-population}, and 18 of the 8662 in
scope of the retention figure. Its call-site axis counts the
remaining 8644 by class; the pooled total it prints, 7758, leaves
out the 886 statement-like call sites, whose row is dropped because
no tool retains any.

%% file: figures/tex/fig-macros-population.tex
\begin{figure}[t]
  \centering
  \includegraphics[width=\linewidth]{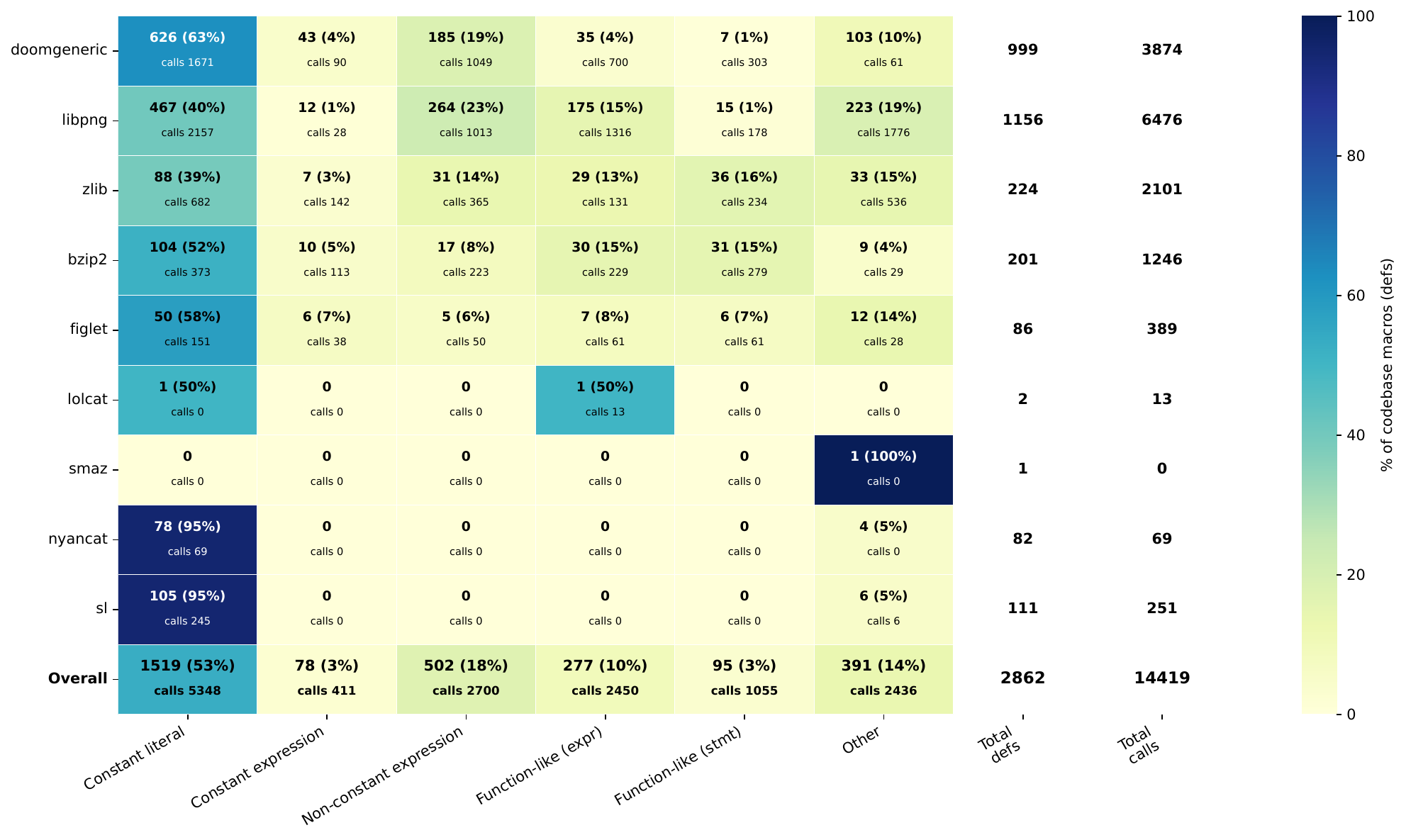}
  \caption{Per-codebase macro population, translator-agnostic. Each
cell shows the number of distinct macro definitions in that class
(top, with the per-row percentage) and the corresponding total
call-site count (bottom). The call-site totals exclude the 19
sites of the two names whose class differs between
conditional-compilation branches (\S\ref{sec:supp-population}).}
  \Description{Per-codebase macro population table.}
  \label{fig:macro-population}
\end{figure}

%% file: body/supplementary_safety.tex
\section{Safety Roadmap}
\label{sec:supp-safety-roadmap}

This section sketches the planned middle-end passes that would
take \sys{}'s structure-preserving output and progressively refine
it toward fully safe, idiomatic Rust. The roadmap is future work
relative to the main paper and is reported here for the
interested reader.

\paragraph{Related work on safety refinement.}
Beyond syntactic translation, several lines of work pursue formal
semantics or refinement passes to map C constructs into safe Rust.
Fromherz and Protzenko~\cite{fromherz_scylla_2026} propose Scylla,
which expresses C's pointer arithmetic with Rust slices; Zhang et
al.~\cite{Zhang_Crown_2023} propose CROWN, a static ownership analysis
that translates C pointers into safe Rust equivalents where ownership
can be inferred; Hong
and Ryu~\cite{hong_automatically_2025} argue for
chaining refinement passes to close the safety and idiomaticity
gap. These approaches improve safety but do not address macro
preservation, which remains a key obstacle to maintainability.

A \emph{migration target} (main paper, \S1) requires both
\emph{structure preservation} (the focus of the main paper) and
\emph{safety refinement}. \sys{}'s IR is designed so that safety
analyses operate over preserved abstractions---macros, helper
traits, and conditional branches (future work)---rather than over the flat
post-preprocessor form that prior tools emit. Our roadmap
incorporates the techniques above as middle-end passes over
\sys{}'s IR.

\paragraph{IR vs. wrapper architectures.}
Where macro awareness lives, inside the translator's IR or in a
wrapper around an existing translator's output, shapes what a later
safety refinement has to work with. The concurrent Hayroll
system~\cite{peng_hayroll_2026} demonstrates that macro and
conditional-compilation awareness can be retrofitted onto an
existing translator through tagging and source-correspondence: a
complementary path to ours. However, Hayroll's wrapper leaves
the underlying translator's rendering of the code as it is and
rebuilds the macro and \texttt{\#[cfg]} structure around it, so
it inherits the \texttt{unsafe} idioms C2Rust emits today. The refinement
passes Hong and Ryu argue for each run a static analysis over
C2Rust's output, owner identification and lifetime inference among
them, and rewrite the code accordingly. Run over C2Rust's output they see macro-expanded Rust and rewrite
every expansion separately; a wrapper that reconstructs macro
definitions from that output afterwards, as Hayroll does, must then
derive one definition from expansions that may no longer agree.
Run over an IR that keeps macro definitions and invocations as
nodes, they can analyse and rewrite a macro once, at its
definition. \sys{}'s IR is
such an IR by construction; a wrapper around a black-box translator
does not expose one. The two approaches are therefore
complementary rather than competing: structure preservation can
be achieved either way; we expect the IR-integrated route to be the
more direct road from structure to safety, though we have not
evaluated this yet.

A third, also-concurrent point in the design space is
Cpp2Rust~\cite{popescu_cpp2rust_2026}, which from a C++ subset
emits Rust that is safe \emph{up front} by routing every variable
through \texttt{Rc<RefCell<\textit{T}>{}>} and modelling pointers
with a runtime type, then iteratively eliminating the wrappers
where an intra-procedural analysis proves them unneeded.
This trades structure preservation and a slowdown of between 2\%
and 6$\times$ on their two applications for the strong guarantee that the generated program contains no
\texttt{unsafe} code outside a fixed runtime library. \sys{}'s
IR-integrated route takes the opposite tradeoff: \texttt{unsafe}
remains in the initial translation, but the structure on which to
recover safety statically (via the passes above) is preserved.
Cpp2Rust shows that the runtime-safety direction is viable for at
least a C++ subset; \sys{} is the platform for the future
refine-from-structure direction on C.

\paragraph{Planned middle-end passes.}
The IR's middle end is currently empty by design. We plan four
classes of pass, each with a narrow responsibility:
\begin{itemize}[leftmargin=*]
\item \textbf{Pointer-provenance analysis.} Track which raw
      pointer values originate from owned allocations, references,
      or external/FFI sources; promote pointers to references or
      smart pointers when provenance permits, in the spirit of
      CROWN~\cite{Zhang_Crown_2023}.
\item \textbf{Alias-set inference.} Group pointers that may alias
      to drive borrow-check-friendly translations; when aliasing
      is provably absent, emit \texttt{\&mut} references rather
      than \texttt{*mut}.
\item \textbf{Cast-kind disambiguation.} \sys{}'s
      \texttt{helper\_cast} macro is dispatched on an explicit
      ``kind''; that taxonomy is also a natural analysis target
      for replacing unchecked casts with typed conversions where
      statically safe.
\item \textbf{Macro-level ownership recognizers.} Recurring macro
      idioms in real codebases (e.g., a macro that takes a
      pointer and stores it in a struct field, or a macro that
      releases a resource) are amenable to pattern-recognizer
      passes that rewrite them into safe Rust idioms
      (\texttt{Box::into\_raw}, \texttt{Drop}, etc.).
\end{itemize}

\paragraph{Classifying provably safe macros.}
A first concrete pass already within reach of the existing
infrastructure is a \emph{trivially-safe macros} classifier. For
each preserved macro definition, the pass answers \emph{yes}
(statically safe to leave as-is in a future safe-Rust profile)
or \emph{don't know} (further analysis required). The criterion
is intentionally conservative: a macro is ``yes'' iff its body
combines its parameters, literals and immutable constants by
arithmetic, comparison, or field projection over \texttt{Copy}
scalar types; takes parameters by value; and calls only functions
that are themselves safe and return such scalars. Anything
else---raw pointers (which are \texttt{Copy} but not safe to
dereference), unions, mutable statics, calls into FFI or into
\texttt{unsafe fn}---is ``don't know.'' The pass is intended
not as a complete safety analysis but as a starting point
for the heavier refinement chain above, by labelling the macros
that contribute zero unsafety.

%% file: body/supplementary_overhead.tex
\section{Performance Overhead}
\label{sec:supp-overhead}

This section reports the per-benchmark performance figures
summarised at the end of \S5.2 of the main paper. The main conclusion there is
that \sys{}'s output runs within 5--10\% of Clang on most
workloads; below we report the full breakdown for runtime and
memory across the two benchmark suites.

\input{figures/tex/fig-overhead-heatmap}

\paragraph{Translators.}
We selected \emph{C2Rust} and \sys{} for the rule-based
comparison. As baselines, we include
\emph{Clang} and \emph{GCC} with aggressive optimization, and a
hand-written idiomatic Rust implementation of each benchmark.
This separates overhead from translation versus differences
inherent to the target language.
Compilation flags followed standard release configurations:
\texttt{-O3 -DNDEBUG} for Clang/GCC, and
\texttt{-C opt-level=3 -C overflow-checks=off -C debuginfo=0}
for Rust.

\paragraph{Benchmarks.}
The study uses 32 single-file C programs in three groups
(Table~\ref{tab:supp-overhead-benchmarks}). Each program takes its
problem size on the command line, chosen so that the Clang build
runs for 1 to 32\,s (median 3.3\,s; the A* search is the exception,
at about 30\,ms), and prints a result checksum.
\emph{Microbenchmarks} (19 programs, 31--99 lines each) isolate one
class of operation each: integer and bit arithmetic, floating
point and the C math library, graph and dynamic-programming
algorithms, and memory and string handling. They were written for
this study with AI assistance and manually curated for correctness
and reproducibility; they stress low-level constructs, not macros
(two of them define a single constant).
\emph{Application-style programs} (6 programs, 20--632 lines) come
from established benchmark collections: binary-trees,
fannkuch-redux, and n-body from the Computer Language Benchmarks
Game~\cite{Measured68:online}, whose headers credit the original
contributors; an A* grid path search adapted from Rosetta
Code~\cite{RosettaC15:online}, with the grid enlarged from
$10\times10$ to $201\times201$; and two small classics, recursive
Fibonacci and a Leibniz-series approximation of~$\pi$.
\emph{Macro-intensive benchmarks} (7 sorting algorithms, 43--72
lines) use macros throughout their core logic: 27 definitions, 18
of them function-like. They span the classes of \S3.5 of the main
paper: constants (e.g., \texttt{MAXVAL}, \texttt{BUCKETS}),
expression-like function macros (e.g., \texttt{BUCKET\_INDEX},
invoked in bucket sort's inner loop), statement macros in
\texttt{do \{ \dots\ \} while (0)} form (e.g., \texttt{SWAP},
\texttt{MERGE}), and loop-header macros (e.g., \texttt{FOR}). By
design, \sys{} preserves the first two classes as Rust macros and
conservatively expands the last two, so this suite measures both
the cost of preserved macros on hot paths and that of
expansion. Hand-written idiomatic Rust versions exist for all
programs except the A* search and the $\pi$ approximation; for
these two, and for \texttt{branch} and bucket sort, whose Rust
versions did not produce a valid measurement, the idiomatic-Rust
cells of Figure~\ref{fig:overhead} are empty.

\begin{table*}[t]
  \centering
  \small
  \begin{tabular}{@{}l l p{0.62\linewidth}@{}}
    \toprule
    Group & Source & Programs \\
    \midrule
    Micro (19) & written for this study & integer/bits: \texttt{int}, \texttt{bitops}, \texttt{modulo}, \texttt{branch}, \texttt{hash} (FNV-1a);
      floating point: \texttt{math}, \texttt{mandelbrot}, \texttt{dft}, \texttt{fft} (radix-2 Cooley--Tukey), \texttt{gausselim}, \texttt{nn} (neural-network forward pass), \texttt{kmeans} (Lloyd);
      graphs/DP: \texttt{dijkstra}, \texttt{floyd} (Floyd--Warshall), \texttt{lcs};
      memory/strings: \texttt{transpose}, \texttt{search} (linear and binary), \texttt{primes} (sieve), \texttt{string} \\
    Application (6) & Benchmarks Game & \texttt{bintrees} (binary-trees), \texttt{fanred} (fannkuch-redux), \texttt{nbod} (n-body) \\
      & Rosetta Code & \texttt{Astar} (A* search on a $201\times201$ grid) \\
      & classic & \texttt{fib} (recursive Fibonacci), \texttt{calcpi} (Leibniz series) \\
    Macro-intensive (7) & written for this study & bucket, counting, insertion, merge, radix, selection, and shell sort \\
    \bottomrule
  \end{tabular}
  \caption{The 32 programs of the performance study, by group and
  source (names as in Figure~\ref{fig:overhead}).}
  \label{tab:supp-overhead-benchmarks}
\end{table*}

\paragraph{Machine.}
All measurements ran on a single host: an Intel Core Ultra 7
165H (Meteor Lake, 16 cores / 22 threads, base
400\,MHz / boost 5.0\,GHz), 30\,GiB RAM, Ubuntu 24.04.3 LTS
(kernel 6.8.0-45-generic), gcc 13.3.0, clang 18.1.8. The Rust
configurations (\sys{}, C2Rust, and idiomatic Rust) were compiled
with the container's then-current nightly rustc (September 2025);
for reproduction, the artifact pins \texttt{nightly-2025-09-25}.

\paragraph{Evaluation protocol.}
The measurement campaign is orchestrated with
benchkit~\cite{benchkit_icpe26}; per-tool build environments are
Docker images composed with pythainer~\cite{pythainer_joss}.
Each benchmark was executed 10 times under controlled conditions.
We measured (i) \emph{runtime}, reported as the median execution
time relative to the \emph{Clang} mean (normalized to 1.0), and
(ii) \emph{memory usage}, collected via Valgrind's
\texttt{massif} tool~\cite{nethercote2007valgrind}. For runtime,
we pinned benchmark processes to a fixed CPU core to reduce
scheduling noise. For memory, we report average consumption
across the program's lifetime, including stack usage.

\paragraph{Results.}
Figure~\ref{fig:overhead} summarizes overheads, reported as the
median of 10 runs normalized to the mean Clang measurement for
both runtime and memory.

Across both microbenchmarks and macro-intensive sorting
benchmarks (Figure~\ref{fig:overhead-runtime}), \sys{} is within 10\% of Clang on 24 of the 32 benchmarks, faster
on six, and more than 10\% slower on two (\texttt{lcs} and
\texttt{fib}). C2Rust has the same profile (25 of 32 within 10\%,
slower on the same two), so macro preservation shows no systematic
runtime difference from C2Rust. Idiomatic Rust spreads more widely
in both directions: 7 of its 28 valid benchmarks are within 10\% of
Clang, 12 are faster and 9 slower, up to 2.2$\times$. These
differences are consistent with language-level choices (bounds
checks, iterator patterns, library calls), but the measurements do
not isolate their causes.

Memory overhead trends follow a comparable pattern
(Figure~\ref{fig:overhead-memory}). \sys{} matches C2Rust almost entirely (median
difference 0.01), so macro preservation shows no measurable memory
cost relative to C2Rust. Against Clang, however, twelve of the 32
benchmarks use 1.7--4.3$\times$ more memory under \sys{} (e.g.,
\texttt{math}, \texttt{mandelbrot}, \texttt{nbod}), and on each of
them C2Rust and, where it has a valid measurement, idiomatic Rust
use a comparable or larger multiple (on \texttt{nbod}, 6.4--6.7$\times$;
idiomatic Rust has no measurement for \texttt{branch} and
\texttt{calcpi}). This is consistent with language and runtime
characteristics rather than with translation artefacts.

%% file: figures/tex/fig-overhead-heatmap.tex
\begin{figure*}[t]
  \centering
  \begin{subfigure}{\textwidth}
    \centering
    \includegraphics[width=\textwidth]{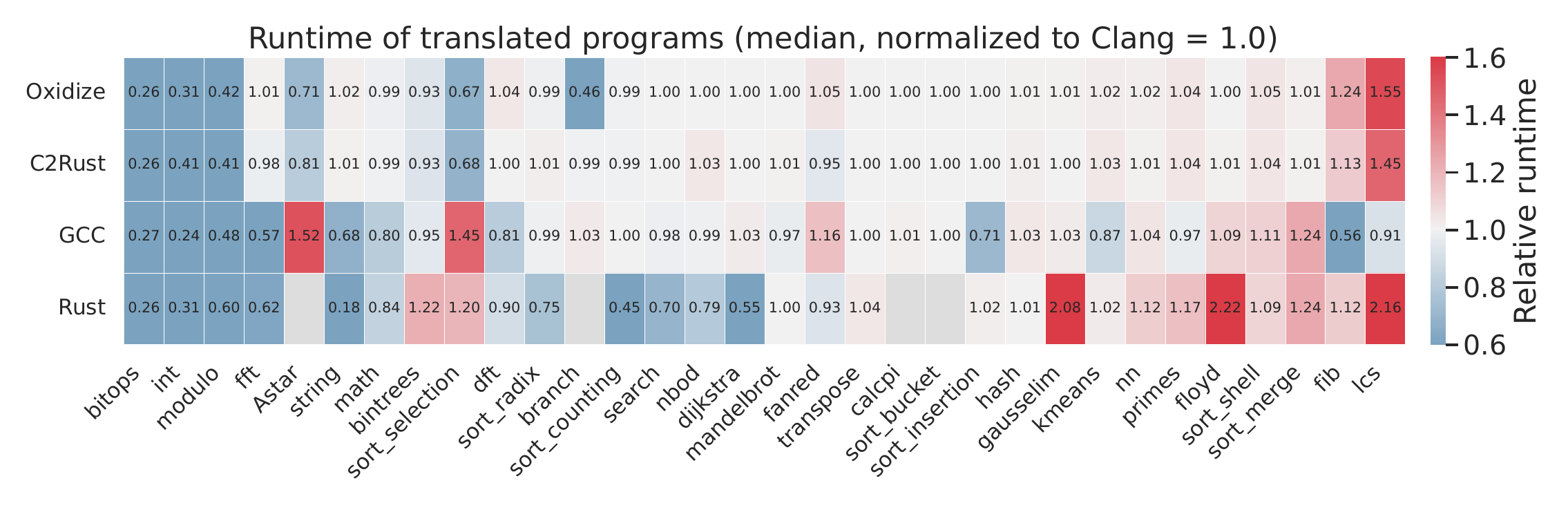}
    \caption{Runtime overhead. \sys{} is within 10\% of \texttt{Clang}
    on 24 of the 32 benchmarks, with the same profile as C2Rust;
    idiomatic Rust spreads more widely in both directions.}
    \Description{Heatmap of runtime relative to Clang. Rows are translation tools, columns are benchmarks, and each cell holds the median runtime normalised to the Clang baseline, blue below 1.0 and red above.}
    \label{fig:overhead-runtime}
  \end{subfigure}

  \vspace{1em}

  \begin{subfigure}{\textwidth}
    \centering
    \includegraphics[width=\textwidth]{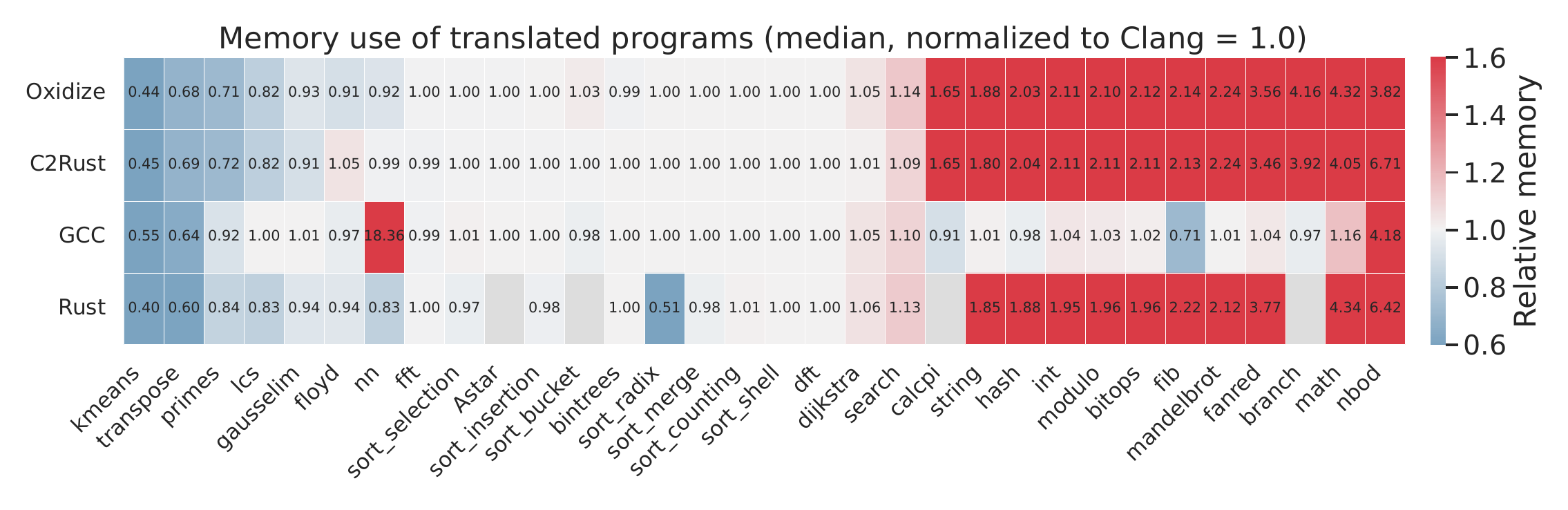}
    \caption{Memory usage overhead (average memory use, heap and stack).
  \sys{} matches C2Rust almost entirely; the twelve benchmarks on
  which it uses 1.7--4.3$\times$ the memory of C (e.g., \texttt{math},
  \texttt{mandelbrot}, \texttt{nbod}) show a comparable or larger
  multiple under C2Rust and, where measured, idiomatic Rust, consistent
  with language and runtime differences rather than translation artifacts.}
    \Description{Heatmap of average memory use (heap and stack) relative to Clang, laid out like the runtime heatmap above: tools as rows, benchmarks as columns, values normalised to the Clang baseline.}
    \label{fig:overhead-memory}
  \end{subfigure}

  \caption{Relative overhead of translators compared to \texttt{Clang} (baseline $=1.0$).
  Each cell shows the median value normalized to \texttt{Clang} for a given benchmark
  (columns) and translation tool (rows). Values below~1.0 (blue) indicate lower
  overhead than \texttt{Clang}, while values above~1.0 (red) indicate higher overhead.}
  \Description{Two stacked heatmaps of overhead relative to Clang, runtime above and memory below, sharing the same benchmark columns and tool rows.}
  \label{fig:overhead}
\end{figure*}

%% file: body/supplementary_prompts.tex
\section{LLM Prompt Template}
\label{sec:supp-prompts}

This section reproduces the prompt template used by the two agentic
LLM-backed translators evaluated in \S5.2 of the main paper (Claude
and Codex; the template is identical for both) and states the full
configuration under which they ran, as fixed by the artifact's
per-tool plugins.

\paragraph{Working directory and prompt.}
Each program is translated in a fresh temporary directory holding
\texttt{input.c} (the C source), \texttt{tests.json} (the program's
reference test cases when the dataset provides them, an empty list
otherwise) and, after the run, the requested output file. The
directory containing the C source is linked into the working directory
as \texttt{project/}, so that headers and sibling files resolve; this
template does not mention it (a variant for whole translation units
does, and asks the agent to treat it as read-only). The template below
is instantiated with those three relative paths and handed to the tool
as its single user prompt.

\paragraph{Models and invocation.}
Both tools reach their model through OpenRouter, with the only secret,
the OpenRouter API key, forwarded into the container at runtime.
\emph{Claude} is the Claude Code command-line tool, run headless as
\begin{Verbatim}[fontsize=\small]
claude --print --output-format text --permission-mode bypassPermissions
       --no-session-persistence --bare --settings <file> "<prompt>"
\end{Verbatim}
with a settings file that sets \texttt{ANTHROPIC\_BASE\_URL} to
OpenRouter and pins every model slot (\texttt{ANTHROPIC\_MODEL} and
the default Opus, Sonnet and sub-agent models) to
\texttt{anthropic/claude-opus-4.7}. \emph{Codex} is the Codex
command-line tool, run as
\begin{Verbatim}[fontsize=\small]
codex --ask-for-approval never exec --sandbox danger-full-access --ephemeral
      --ignore-rules --skip-git-repo-check --cd <workdir> - < prompt.txt
\end{Verbatim}
with the prompt on standard input, inside a
freshly initialised Git repository (Codex expects one); its
configuration file selects provider \texttt{openrouter}, model
\texttt{openai/gpt-5.3-codex} and \texttt{model\_reasoning\_effort =
high}. Everything else is at each tool's defaults.

\paragraph{Bounds and outcome.}
No turn or iteration limit is passed; each tool's own compile-and-fix
loop is bounded only by the per-program cap that applies to every
translator, \texttt{transpile\_timeout\_s = 600} in each dataset's
configuration, enforced by the pipeline killing the tool's process.
A run that exceeds the cap, exits non-zero, or leaves no file at the
requested output path counts as a translation failure; otherwise the
file is compiled, run and compared exactly like every other tool's
output. Each program was run once; the tools are non-deterministic.

\paragraph{C2SaferRust.}
The third LLM-backed tool is not agentic and takes no prompt from us.
It runs C2Rust (v0.22.1, the same pin as the C2Rust column) and then
C2SaferRust's own repair pipeline at its pinned commit
(\texttt{e8d5c3a1}), which calls \texttt{gpt-4o-mini} (the
2024-07-18 snapshot, hard-coded upstream) through the OpenAI API
redirected to OpenRouter, under the same 600\,s cap.

\paragraph{The template.}

\begin{Verbatim}[breaklines=true,fontsize=\small]
You are a C-to-Rust transpiler.

Translate the following complete C source file to a complete standalone Rust
2021 program that can be compiled directly with rustc.

Requirements:
- Preserve the observable behavior of the original program, including stdout, stderr, command-line argument handling, and exit status.
- Use only the Rust standard library.
- Prefer safe Rust, but use unsafe blocks if absolutely needed to preserve the C semantics.
- You may create temporary files and run local commands inside the working directory to check your translation.
- Try to compile and run the generated Rust code to check the correctness
- The original C source is available at: {source_path}
- Unit test cases are available at: {tests_path}
- Do not modify the C source code or the test cases.
- Write the final Rust source to exactly this path: {output_path}
- Do not write outside the working directory.
- Do not explain the translation.
\end{Verbatim}